\documentclass[twocolumn,prl,preprintnumbers,aps,10pt]{revtex4-2}
\usepackage{hyperref}
\usepackage[utf8]{inputenc}
\usepackage{amsmath} 
\usepackage{amsfonts} 
\usepackage{amssymb}
\usepackage{graphicx} 
\usepackage{color}
\usepackage{mathtools}
\usepackage[italicdiff]{physics}

\newcommand{\ri}{\mathrm{i}}
\newcommand{\Gc}{G_\mathrm{C}}
\newcommand{\xic}{\xi_\mathrm{C}}
\newcommand{\xiQ}{\xi_{q}}
\begin{document}
\preprint{KEK-TH-2456, J-PARC-TH-0278}
\title{Emergent higher-form symmetry in Higgs phases with superfluidity}
\author{Yoshimasa Hidaka}
\email{hidaka@post.kek.jp}
\affiliation{
KEK Theory Center, Tsukuba 305-0801, Japan
}
\affiliation{
Graduate University for Advanced Studies (Sokendai), Tsukuba 305-0801, Japan
}
\affiliation{
RIKEN iTHEMS, RIKEN, Wako 351-0198, Japan
}
\affiliation{
  Department of Physics, Faculty of Science, University of Tokyo, 7-3-1 Hongo Bunkyo-ku Tokyo 113-0033, Japan
}

\author{Dan Kondo}
\email{dan.kondo@ipmu.jp}
\affiliation{
Kavli Institute for the Physics and Mathematics of the Universe (WPI), University of Tokyo
Institutes for Advanced Study, University of Tokyo, Kashiwa 277-8583, Japan
}
\begin{abstract}
We address emergent higher-form symmetry in Higgs phases with superfluidity. The emergent symmetry appears if a matter field is invariant under a transformation of a common subgroup of gauge and global $\mathrm{U}(1)$ symmetries. We explicitly construct the symmetry generator that is topological and gauge invariant in a low-energy effective theory. Such emergent symmetry is helpful to distinguish the Higgs and other phases with superfluidity, such as phases of QCD at finite density.
We also discuss the possibility of phase transition between these phases.
\end{abstract}
\maketitle

\paragraph{Introduction}
Classification of phases of matter is one of the fundamental problems in many-body physics.
In particular, understanding what kind of phases exist in gauge theories is of great interest for understanding our universe described by the standard model as a gauge theory.
Typical phases in gauge theory are Coulomb, confinement, and Higgs (superconducting) phases, which are not always separated by phase transitions~\cite{tHooft:1979yoe,Osterwalder:1977pc,Fradkin:1978dv,Banks:1979fi}. The question is, what conditions distinguish these phases? 

In QED in $(3+1)$ dimensions, there are distinguishable Coulomb and Higgs phases.
The Higgs phase is characterized by the Bose-Einstein condensation of charged particles.
The conventional order parameter is the expectation value of a charged field, which is often referred to as the spontaneous breaking of gauge symmetry.
However, the order parameter is not gauge invariant; thus, it is not a physical observable.
Although the gauge invariant part of the order parameter is physical, it does not break the gauge symmetry by definition~\cite{tHooft:1979yoe}. In this sense, the local field is not useful to distinguish these phases.
From the modern point of view, the Higgs phase is rather characterized by a nonlocal order parameter, 't Hooft loop in the case of QED~\cite{Tong,Gukov:2013zka}.

Higher-form symmetry is helpful to characterize such phases of gauge theory as global symmetry~\cite{Banks:2010zn,Kapustin:2014gua,Gaiotto:2014kfa}.
Ordinary global symmetry is symmetry for a local field (charged object) whose transformation is generated by a codimension-one topological and gauge invariant object. The ``topological'' means that correlation functions are invariant under deformation of codimension-one subspace unless it crosses a charged object.
This concept can be generalized to symmetry for $p$-dimensional extended objects, which is the higher-form symmetry. The symmetry transformation is generated by a codimension-$(p+1)$ dimensional topological and gauge invariant object.

In gauge theory, there are several extended objects. The fundamental ones in $(3+1)$ dimensions are the Wilson and 't Hooft loops, which are one-dimensional objects.
The Higgs and Coulomb phases in QED can be classified by the magnetic one-form $\mathrm{U}(1)$ symmetry corresponding to the conservation of magnetic fluxes,
whose charged objects are the 't Hooft loops. 
In the Coulomb phase, the magnetic symmetry is spontaneously broken, while the Higgs phase is not.
The unbroken magnetic $1$-form symmetry means that the 't Hooft loop as the order parameter obeys the area law, reflecting that the magnetic flux is confined in the Higgs phase~\cite{Tong}.

If a charge $q\,(\neq\pm1)$ particle is condensed in QED, the resultant low-energy effective theory is the $\mathbb{Z}_q$ gauge theory, which exhibits topological order~\cite{Hansson:2004wca,Wen:2016ddy}. In this case, there are emergent higher-form symmetries, which are $\mathbb{Z}_q$ electric one-form and $\mathbb{Z}_q$ magnetic two-form symmetries. These emergent symmetries characterize the $\mathbb{Z}_q$ valued magnetic flux inside vortices.
In a superconductor, the cooper pair has the charge $q=-2$. 
Therefore, the vortex has a $\pi$-flux.

The characterization of Higgs phases of nonabelian gauge theory is more subtle. For a $\mathrm{SU}(N)$ gauge theory, there is no magnetic one-form symmetry.
If the gauge field couples to a fundamental representation matter, there is no electric one-form symmetry. If the charged field condenses and completely breaks the gauge symmetry in the conventional sense, no global symmetry distinguishes the confinement and Higgs phases; thus, there need not be a phase transition between the two phases~\cite{Fradkin:1978dv,Banks:1979fi}.

QCD with three flavors at finite density is a similar situation. 
It is expected that the confinement phase is realized in the low-density hadronic matter and the Higgs phase called the color-flavor locked phase in the high-density quark matter (See, e.g., Ref.~\cite{Alford:2007xm} for a review). 
In both phases, $\mathrm{U}(1)$ baryon and chiral symmetries are spontaneously broken. 
The global symmetry-breaking pattern is the same between these phases. 
From these observations, the continuity between the hadronic and quark matter phases is conjectured by Sch\"{a}fer and Wilczek~\cite{Schafer:1998ef},
and the spectrum matching between hadronic phases and color superconducting phases was discussed~\cite{Hatsuda:2006ps}, including vortex excitations for three flavors~\cite{Alford:2018mqj,Chatterjee:2018nxe}, and two flavors~\cite{Fujimoto:2019sxg,Fujimoto:2020dsa,Fujimoto:2021wsr,Fujimoto:2021bes}.

One particular feature in QCD with three flavors at finite density is that both confinement and Higgs phases exhibit superfluidity.
In a superfluid phase, there is an emergent $\mathrm{U}(1)$ two-form symmetry~\cite{Delacretaz:2019brr}, corresponding to a worldsurface of a vortex as a charged object. The charge is an integer winding number.
In addition, in the color flavor-locked phase, there is another type of vortex called the ``nonabelian vortex'', which has a fractional winding number~\cite{Balachandran:2005ev,Nakano:2007dr} (See Ref.~\cite{Eto:2013hoa} for a review). 
In addition, the nonabelian vortex generates a nontrivial $\mathbb{Z}_3$ valued Aharonov–Bohm phase (holonomy)~\cite{Cherman:2018jir}.
If it is in a gapped phase, such an Aharonov–Bohm phase is the signal of the topological order, but it is not always the case in a gapless phase~\cite{Hirono:2018fjr,Hirono:2019oup}.
Nevertheless, the holonomy is useful to characterize the Higgs phase with superfluidity~\cite{Cherman:2020hbe}. 

In this letter, we address the emergent higher-form symmetry that distinguishes the Higgs and other phases with superfluidity, using a low-energy effective theory.
By explicitly constructing the topological objects, we clarify the mechanism of the emergence of higher-form symmetry.
Our emergent symmetry is a related but different symmetry from the holonomy.
It is shown that $\mathbb{Z}_3$ valued holonomy in QCD is not stable under an explicit breaking of flavor symmetry~\cite{Hirono:2018fjr}.
In contrast, our higher-form symmetry is robust against the explicit breaking term.

\paragraph{Formulation and result}
We are interested in a model with a gauge symmetry $\Gc$ and a global $\mathrm{U}(1)_\mathrm{Q}$ symmetry in $(d+1)$ spacetime dimensions. Here, $d$ represents the spatial dimension.
Throughout this paper, we use subscripts ``$\mathrm{C}$'' (color) and ``$\mathrm{Q}$'' (quark) for gauge group and global $\mathrm{U}(1)$ group, respectively.
Let $\phi_\ell$ be a matter field with a representation $(\rho_\ell, q_\ell)$ of $\Gc\times \mathrm{U}(1)_\mathrm{Q}$. The label $\ell$ distinguishes different types of matter fields.
The field transforms under $\Gc\times \mathrm{U}(1)_\mathrm{Q}\ni (g,e^{\ri\theta})$ as $\phi_\ell\to e^{\ri q_\ell\theta}{\rho_\ell}(g)\phi_\ell$, where ${\rho_\ell}(g)$ is the representation matrix of the gauge group.
We assume that, except the identity, there is no element $g\in\Gc$ such that ${\rho_\ell}(g)=1$ for all $\ell$.
In this case, the model has no electric one-form symmetry.

Suppose there exists elements $(\gamma, e^{\ri\omega} )\in \Gc\times\mathrm{U}(1)_\mathrm{Q}$ such that 
$e^{\ri q_\ell\omega}{\rho_\ell}(\gamma)=1$ for all $\ell$ whose number of elements is finite $K$.
Then, the set of the elements forms $\mathbb{Z}_K$ group, which is a common subgroup of $\Gc$ and $\mathrm{U}(1)_\mathrm{Q}$.
This $\mathbb{Z}_K$ is the redundancy of the transformation.
The faithful transformation of matter fields is generated by $(\Gc\times\mathrm{U}(1)_\mathrm{Q})/\mathbb{Z}_K$.
Therefore, the global $\mathrm{U}(1)$ symmetry is not $\mathrm{U}(1)_\mathrm{Q}$ but $\mathrm{U}(1)_\mathrm{Q}/\mathbb{Z}_K\eqqcolon \mathrm{U}(1)_\mathrm{B}$. In the case of QCD, the matter fields are quarks with the fundamental representation. The center $\gamma=e^{\ri2\pi n/3}\in\mathrm{SU}(3)_\mathrm{C}$ and $e^{-\ri2\pi n/3}\in\mathrm{U}(1)_\mathrm{Q}$ ($n\in\mathbb{Z}$) keeps the transformation of quarks invariant. Therefore, $K=3$, and $\mathrm{U}(1)_\mathrm{B} = \mathrm{U}(1)_\mathrm{Q}/\mathbb{Z}_3$ is nothing but the $\mathrm{U}(1)$ baryon symmetry. In general, matter fields may have another symmetry, like flavor symmetry in QCD, which does not play an important role in our argument, so we only focus on the gauge symmetry and global $\mathrm{U}(1)$ symmetry.

We are interested in confinement and Higgs phases when the global symmetry $\mathrm{U}(1)_\mathrm{B} = \mathrm{U}(1)_\mathrm{Q}/\mathbb{Z}_K$ is spontaneously broken into its subgroup. We assume that excitations are all gapped except the Nambu-Goldstone modes associated with the spontaneous breaking of $\mathrm{U}(1)_\mathrm{B}$ symmetry.
Although the confinement is not well defined precisely because there is no electric nor magnetic one-form symmetry, we will refer to the phase contrasted with the Higgs phase as the confinement phase.

Let $\Phi$ be a field with a representation $(\rho_{\Phi},q)$ under $\Gc\times\mathrm{U}(1)_\mathrm{Q}$ that is supposed to condense in the Higgs phase.
$\Phi$ may be either an elementary or a composite field of $\phi_\ell$. 
It transforms under gauge and global symmetries as
$\Phi\to e^{\ri q\theta}{\rho_{\Phi}}(g)\Phi$.
Let $O(\Phi)$ be a gauge invariant combination of $\Phi$ that has $(1,Kq)$, where $1$ represents the trivial representation.
We consider the case $\expval{O(\Phi)}\neq0$ in both confinement and Higgs phases, in which the global $\mathrm{U}(1)_\mathrm{B}$ symmetry is spontaneously broken into $\mathbb{Z}_{q}$ symmetry.

We would like to distinguish phases with the condensations, $O(\Phi)$ and $\Phi$.
Since $\Phi$ is not a gauge invariant field, we cannot use itself as the order parameter. Instead, we will show that  $\mathbb{Z}_{K}$ $(d-1)$-form symmetry emerges in a low-energy effective theory in the $\Phi$ condensed phase,
where the common subgroup $\mathbb{Z}_{K}$ of $\Gc$ and $\mathrm{U}(1)_\mathrm{Q}$ plays the essential role.

Let us start with the confinement phase. The coset space of
$\mathrm{U}(1)_\mathrm{B}$ symmetry breaking is $\mathrm{U}(1)_\mathrm{B}/\mathbb{Z}_{q}\simeq \mathrm{U}(1)$, and the low-energy effective Lagrangian is constructed by
the Mauler-Cartan one-form,
$\ri\xi_\mathrm{B}\dd\xi_\mathrm{B}^\dag=\dd{\varphi_\mathrm{B}}=\partial_\mu\varphi_\mathrm{B} \dd{x^\mu}$, where $\xi_\mathrm{B}=e^{\ri\varphi_\mathrm{B}}\in \mathrm{U}(1)_\mathrm{B}/\mathbb{Z}_{q}$.
In this case, there is an emergent $\mathrm{U}(1)$ $(d-1)$-form symmetry, whose symmetry generator is
\begin{equation}
    U(e^{\ri\theta_\mathrm{B}},C) \coloneqq \exp(\ri\frac{\theta_\mathrm{B}}{2\pi}\int_C\dd{\varphi_\mathrm{B}}),
    \label{eq:toopological_object_U(1)B}
\end{equation}
where $C$ is a closed line.
$U(e^{\ri\theta_\mathrm{B}},C)$ is topological because it is invariant under a deformation $C\to C+\delta C$. This can be shown as follows. Noting $\delta C$ can be expressed by the boundary of a one-dimensional higher subspace $S$, $\delta C=\partial S$,
we find $\int_{C+\delta C}\dd{\varphi}=\int_{C}\dd{\varphi}+\int_{\partial S}\dd{\varphi}=\int_{C}\dd{\varphi}+\int_S\dd(\dd\varphi) =\int_{C}\dd{\varphi}$.
Here, we employed Stokes' theorem and $\dd(\dd\varphi)=0$.
The parameter $e^{\ri\theta_\mathrm{B}}$ is $\mathrm{U}(1)$ valued, and $C$ is the codimension-$d$ ($=$ dimension-one) subspace, so that this is the $\mathrm{U}(1)$ $(d-1)$-form symmetry. In a superfluid phase, there are vortex excitations, which are heavy, so that the creation of vortices is suppressed at low energy. The emergence of the higher form symmetry is caused by the absence of dynamical vortices at low energy. The charged object is the worldsurface of a superfluid vortex as a defect.

Let us move on to the Higgs phase. We focus on the case that the symmetry breaking pattern, including the gauge group, is 
\begin{equation}
    \frac{\Gc\times \mathrm{U}(1)_\mathrm{Q}}{\mathbb{Z}_K}\to
        \mathbb{Z}_q, 
\end{equation}
and the coset space is
\begin{equation}
    \frac{\Gc\times \mathrm{U}(1)_\mathrm{Q}}{\mathbb{Z}_K\times \mathbb{Z}_{q}}
    =\frac{\Gc\times \mathrm{U}(1)_q}{\mathbb{Z}_K}.
\end{equation}
Here, we defined $\mathrm{U}(1)_q\coloneqq\mathrm{U}(1)_{\mathrm{Q}}/\mathbb{Z}_q$.
The essential point is that the common part of $\Gc\times \mathrm{U}(1)_q$ is divided by $\mathbb{Z}_K$, 
which will be the origin of the emergent symmetry.

If we choose an element $(\xic, \xiQ)=(e^{\ri\varphi_\mathrm{C}},e^{\ri \varphi_q})\in\qty(\Gc\times \mathrm{U}(1)_q)/\mathbb{Z}_K$, it transforms as
    $(\xic, \xiQ)\to (g \xic h^{\dag}, e^{\ri\theta}\xiQ h^{\dag})$,
under $(\Gc,\mathrm{U}(1)_q)\ni (g,e^{\ri\theta})$, where $h\in \mathbb{Z}_K$. In general, $h$ depends on $\xic$ and $\xiQ$.
We note $h^K=1$ for an arbitrary $h\in \mathbb{Z}_K$; thus  $\xiQ^K$ has charge $Kq$ and gauge invariant.
By using $\ri \xi_{q}^K \dd (\xi_{q}^K)^{\dag} =K\dd\varphi_q$,
we can construct a topological and gauge invariant object as
\begin{equation}
    U(e^{\ri\theta},C) 
    \coloneqq \exp(\ri \frac{\theta}{2\pi} K\int_C \dd\varphi_q).
\end{equation}
Since $e^{\ri\theta}$ is $\mathrm{U}(1)$ valued
and $K\int_C\dd\varphi_q\in2\pi\mathbb{Z}$, $U(e^{\ri\theta},C) $ is the generator of the $\mathrm{U}(1)$ $(d-1)$-form symmetry, which corresponds to Eq.~\eqref{eq:toopological_object_U(1)B} by $\dd{\varphi_\mathrm{B}}=K\dd{\varphi_q}$.

Let us now show there is another topological object in the Higgs phase.
For this purpose, we introduce
\begin{equation}
    a \coloneqq  \ri\xic \dd \xic^\dag,
\end{equation}
which transforms under a gauge transformation as
\begin{equation}
    a \to g a g^\dag +\ri  g \dd g^\dag + \ri h^\dag\dd{h}.
    \label{eq:gauge_transformation}
\end{equation}
Since $a$ is ``pure gauge'', the field strength vanishes
\begin{equation}
    f\coloneqq \dd a + a\wedge a = 0.
    \label{eq:flat}
\end{equation}

We here define
\begin{equation}
   W(\rho,C)\coloneqq\frac{1}{d_\rho}\tr\rho\qty[ \mathrm{P} \exp(\ri\int_C a)],
   \label{eq:Wilson_loop}
\end{equation}
and would like to show that $W(\rho,C)$ is the generator of emergent $\mathbb{Z}_K$ $(d-1)$-form symmetry.
Here, $\mathrm{P}$ is the path-ordering operator, and $d_\rho$ is the dimension of the representation $\rho$.
By construction, this object is invariant under Eq.~\eqref{eq:gauge_transformation} like Wilson loops.
At the same time, this is topological: $W(\rho,C)$ is invariant under infinitesimal deformation $C\to C+\delta C$ with $\delta C=\partial S$.
This is because the difference is proportional to the curvature on $S$~\cite{Ambrose}, and it vanishes due to the flatness~\eqref{eq:flat}.
$W(\rho,C)$ measures the winding phase of nonabelian vortices.
For $h\in \mathbb{Z}_K$, $(\xic,\xi_q)$ and $(\xic h,\xi_q h)$ are the same point in the coset space.
Therefore, the nonabelian vortices are characterized by $h$, and 
$W(\rho,C)$ takes the value
\begin{equation}
    W(\rho,C)=\frac{1}{d_\rho}\tr\rho(h)=\exp(\ri \frac{2\pi nm_\rho}{K}),
    \label{eq:WZKvalued}
\end{equation}
 where $h=e^{\ri 2\pi n/K}$ ($n\in\mathbb{Z}$) and $m_\rho$ is an integer depending on the representation. For a $\Gc=\mathrm{SU}(N)$ gauge theory coupled with a fundamental representation matter as an example, $K=N$ and $m_\rho$ is equal to the number of boxes in the Young tableau of the representation $\rho$.

The product of $W(\rho_a,C)$ and $W(\rho_b,C)$
is formally expanded by the tensor product of representations as $W(\rho_a,C)W(\rho_b,C)=\sum_{\rho_c}
N_{\rho_a\rho_b}^{\rho_c}W(\rho_c,C)d_{\rho_c}/(d_{\rho_a} d_{\rho_b})$, where $N_{\rho_a\rho_b}^{\rho_c}$ are the tensor-product coefficients, which are nonnegative integers. The factor $d_{\rho_c}/(d_{\rho_a} d_{\rho_b})$ comes from the normalization~\eqref{eq:Wilson_loop}.
At first glance, this product does not satisfy the group rule. However, noting 
$W(\rho_c,C)=W(\rho_d,C)$ if $N_{\rho_a\rho_b}^{\rho_c}\neq0$ and $N_{\rho_a\rho_b}^{\rho_d}\neq0$ in our situation~\eqref{eq:WZKvalued},
we have $W(\rho_a,C)W(\rho_b,C)=W(\rho_c,C)$,
where $\rho_c$ is a representation satisfying $m_c=m_a+m_b\mod K$.
Therefore, $W(\rho,C)$'s satisfy the group rule of $\mathbb{Z}_K$. The emergence of $\mathbb{Z}_K$ $(d-1)$ form symmetry is our main result.
\paragraph{Examples}
Let us show the emergent global symmetry in two examples: $\mathrm{U}(1)_\mathrm{C}\times \mathrm{U}(1)_\mathrm{Q}$ model in $(2+1)$ dimensions and three flavor QCD at high density in $(3+1)$ dimensions.

Let us first consider the $\mathrm{U}(1)_\mathrm{C}\times \mathrm{U}(1)_\mathrm{Q}$ model in $(2+1)$ dimensions discussed in Ref.~\cite{Cherman:2020hbe}, whose Lagrangian has the form,
\begin{align}
    \mathcal{L}=&-\frac{1}{4e^2}F_{\mu\nu}F^{\mu\nu} - V_\mathrm{m}(\sigma)
    -|\partial_\mu\phi_0|^2-m_0^2|\phi_0|^2 \notag\\
    &
    -\lambda_0|\phi|^4
    -|D_\mu\phi_+|^2-|D_\mu\phi_-|^2
    - m^2(|\phi_+|^2+|\phi_-|^2) \nonumber\\
    &- \lambda(|\phi_+|^4+|\phi_-|^4)
    + \epsilon(\phi_+\phi_-\phi_0+\phi_+^*\phi_-^*\phi_0^*)
    .
\end{align}
Here, $m_0$, $m$ are masses of scalar fields; $e^2$, $\lambda_0$, $\lambda$ and, $\epsilon$ are coupling constants.
$F_{\mu\nu}=\partial_\mu A_\nu-\partial_\nu A_\mu$ is the field strength with the gauge field $A_\mu$. The dual photon field $\sigma$  and the potential $V_\mathrm{m}(\sigma)$ are introduced to make the system confined by the Polyakov mechanism~\cite{Polyakov:1976fu}.
The explicit functional form of the potential is not important in our argument.
The dual photon  $\sigma$ is related to
the field strength by $F^{\mu\nu}=\ri e^2\epsilon^{\mu\nu\rho}\partial_\rho\sigma/(2\pi)$, and it explicitly breaks the magnetic zero-form symmetry, $\epsilon^{\mu\nu\rho}\partial_\mu F_{\nu\rho}\propto \partial_\lambda^2\sigma \neq0$.
$\phi_\pm$ and $\phi_0$ are complex scalar fields. $\phi_\pm$ have gauge charges with $\pm1$, while $\phi_0$ is neutral. $\phi_\pm$ couple with the gauge field through the covariant derivative, $D_\mu\phi_\pm = (\partial_\mu\mp\ri A_\mu)\phi_\pm$.
As a global $\mathrm{U}(1)_\mathrm{Q}$ symmetry, 
charge $-1$ is assigned to $\phi_\pm$, while $+2$ is assigned to $\phi_0$.
The coupling $\epsilon \phi_+\phi_-\phi_0$ ensures that there is no other symmetry generated by phase rotations of $\phi_\pm$ and $\phi_0$.

The complex fields transform under $(e^{\ri\lambda},e^{\ri\theta})\in \mathrm{U}(1)_\mathrm{C}\times \mathrm{U}(1)_\mathrm{Q}$ as
    $\phi_0\to e^{2\ri\theta}\phi_0$,
    $\phi_\pm\to e^{\pm\ri\lambda-\ri\theta}\phi_\pm$.
$\phi_0$, $\phi_\pm$ is invariant if $\theta\in (2\mathbb{Z}+1)\pi$ and $\lambda\in (2\mathbb{Z}+1)\pi$.
Therefore, there is $\mathbb{Z}_2$ redundancy, and the transformation faithfully acting on fields is $( \mathrm{U}(1)_\mathrm{C}\times \mathrm{U}(1)_\mathrm{Q})/(\mathbb{Z}_K)_{\mathrm{C}+\mathrm{Q}}$ with $K=2$. This model also has the charge conjugation symmetry $(\mathbb{Z}_2)_\mathrm{charge}:\phi_0\to \phi_0^*$, $\phi_\pm\to \phi_\pm^*$, $A_\mu\to -A_\mu$, and the charged field permutation symmetry, $(\mathbb{Z}_2)_\mathrm{F}:\phi_\pm\to \phi_\mp$, $A_\mu\to -A_\mu$.
Therefore, the global internal symmetry in this model is
\begin{equation}
    \frac{\qty[\mathrm{U}(1)_\mathrm{Q}\rtimes(\mathbb{Z}_2)_\mathrm{charge}]\times (\mathbb{Z}_2)_\mathrm{F}}{(\mathbb{Z}_2)_{\mathrm{C}+\mathrm{Q}}}.
\end{equation}
Here, $\rtimes$ is the semidirect product, which represents the fact that the charge conjugation acts on the $\mathrm{U}(1)_\mathrm{Q}$ symmetry.

We consider two limiting situations. One is $m_0^2\ll -e^4$ and $m^2\gg e^4$, where the system is confined and the $\mathrm{U}(1)_\mathrm{Q}$  global symmetry is spontaneously broken. The other is  $m_0^2\ll -e^4$ and $m^2\ll -e^4$, where the system is in a Higgs phase and the $\mathrm{U}(1)_\mathrm{Q}$  global symmetry is also spontaneously broken.
In both cases, the $\mathrm{U}(1)_\mathrm{Q}$ symmetry is spontaneously broken, so that there is an emergent $\mathrm{U}(1)$ one-form symmetry. In addition, we will show the emergent $\mathbb{Z}_2$ two-form symmetry in the Higgs phase. To obtain a low-energy effective Lagrangian, we set the vacuum expectation value by $\expval{\phi_0}\eqqcolon v_0$, and parametrize the field as $\phi_0=e^{\ri\varphi_0}(v_0+H_0)$.

In the case of $m^2\gg e^4$, charged fields $\phi_\pm$ and neutral field $H_0$ are massive. Their degrees of freedom can be integrated out, and we obtain the low-energy effective Lagrangian as
\begin{align}
    \mathcal{L}_{\text{eff}}=
        -\frac{1}{4e^2}F_{\mu\nu}F^{\mu\nu}-V_\mathrm{m}(\sigma)    
        -\frac{v_0^2}{2}(\partial_\mu\varphi_0)^2+\cdots
    ,
\end{align}
where $\cdots$ represents the higher-order derivative terms.
As mentioned above, there is a $\mathrm{U}(1)$ one-form symmetry, whose symmetry generator is given by
\begin{align}
    U(e^{\ri\theta_0},C)=\exp(\ri\frac{\theta_0}{2\pi}\int_C \dd\varphi_0),
    \label{eq:1-form_operator_U(1)xU(1)}
\end{align}
and the charged object is a worldsurface of $\mathrm{U}(1)$ vortex.

In the case of $m^2\ll -e^4$, charged fields $\phi_\pm$ are also condensed, $\expval{|\phi_\pm|}=v$ in addition to $\expval{\phi_0}=v_0$. Parametrizing charged fields as $\phi_\pm=e^{\ri\varphi_\pm}(v+H_\pm)$, and integrated out massive fields $H_0$ and $H_\pm$, we obtain the effective Lagrangian,
\begin{align}
    \mathcal{L}_{\text{eff}}&=
    -\frac{1}{4e^2}F_{\mu\nu}F^{\mu\nu} - V_\mathrm{m}(\sigma)
    -\frac{v^2}{2}(|D_\mu\varphi_+|^2+|D_\mu\varphi_-|^2)\notag\\
&\quad    -\frac{v_0^2}{2}(\partial_\mu\varphi_0)^2
    +2\epsilon v_0v^2\cos(\varphi_0+\varphi_++\varphi_-)+\cdots
    ,
\end{align}
where $D_\mu\varphi_\pm=\partial_\mu\varphi_\pm\mp A_\mu$.
Note that the last term gives a mass term, and one combination of $\varphi_0$, $\varphi_+$, and $\varphi_-$ is massive.

We would like to find gauge invariant topological objects using $\varphi_\pm$. $\varphi_\pm$ transforms under $\mathrm{U}(1)_\mathrm{C}\times \mathrm{U}(1)_\mathrm{Q}$ as $\varphi_\pm \to \varphi_\pm \pm\lambda -\theta$.
Obviously, the sum, $\varphi_++\varphi_-$, is gauge invariant.
Therefore, 
\begin{align}
U(e^{\ri\theta_+},C)=\exp(-\ri\frac{\theta_+}{2\pi}\int_C \qty(\dd\varphi_++\dd\varphi_-))
\end{align}
is gauge invariant and topological. The parameter $e^{\ri\theta_+}$ is $\mathrm{U}(1)$ valued, so that $U(e^{\ri\theta_+},C)$ is the generator for $\mathrm{U}(1)$ one-form symmetry, which has the same quantum number as Eq.~\eqref{eq:1-form_operator_U(1)xU(1)}.
Similarly, consider
\begin{align}
    W(e^{\ri\theta_-},C)=\exp(\ri\frac{\theta_-}{2\pi}\int_C \qty(\dd\varphi_+-\dd\varphi_-)),
\end{align}
which is topological, but not gauge invariant for general $\theta_-$ because $W(e^{\ri\theta_-},C)$ transforms as
\begin{equation}
    W(e^{\ri\theta_-},C)\to W(e^{\ri\theta_-},C)
    \exp(\ri\frac{\theta_-}{2\pi}2\int_C \dd\lambda).
\end{equation}
Noting $\int_C \dd\lambda\in 2\pi\mathbb{Z}$, and choosing $\theta_-\in \pi \mathbb{Z}$, we can construct a gauge invariant object $W(n,C)$, $n\in \{1,-1\}=\mathbb{Z}_2$, which is the generator of $\mathbb{Z}_2$ one-form symmetry.
 If $n=-1$ and the vortex defect exists with the winding
 $\int (\dd\varphi_+-\dd\varphi_-)/2\in (2\mathbb{Z}+1)\pi$, $W(n,C)$ takes the value of $-1$.

 We have constructed symmetry generators using the explicit model.
We can also construct the symmetry generators using coset variables determined by the symmetry-breaking pattern,
\begin{align}
G&= \frac{ \qty[(\mathrm{U}(1)_\mathrm{C}\times\mathrm{U}(1)_\mathrm{Q})\rtimes(\mathbb{Z}_2)_\mathrm{charge}]\rtimes_{\mathrm{C}} (\mathbb{Z}_2)_\mathrm{F}}{(\mathbb{Z}_2)_{\mathrm{C}+\mathrm{Q}}}\notag\\
& \to
 (\mathbb{Z}_2)_\mathrm{charge}\times (\mathbb{Z}_2)_\mathrm{F}
\eqqcolon H.
\end{align}
 Here, $\rtimes$ presents the action of $(\mathbb{Z}_2)_\mathrm{charge}$ on both $\mathrm{U}(1)_\mathrm{C}$ and $\mathrm{U}(1)_\mathrm{Q}$, while $\rtimes_\mathrm{C}$ presents the action of $(\mathbb{Z}_2)_\mathrm{F}$ on $\mathrm{U}(1)_\mathrm{C}$.
The coset space is
\begin{equation}
    G/H \simeq \frac{\mathrm{U}(1)_\mathrm{C}\times \mathrm{U}(1)_\mathrm{Q}}{(\mathbb{Z}_2)_{\mathrm{C}+\mathrm{Q}} }.
\end{equation}
The degrees of freedom $(\xic,\xiQ)= (e^{\ri \varphi_\mathrm{C}},e^{\ri \varphi_{\mathrm{Q}}})$
transforms under $ (\mathrm{U}(1)_\mathrm{C}\times \mathrm{U}(1)_\mathrm{Q})/(\mathbb{Z}_2)_{\mathrm{C}+\mathrm{Q}}$ as $(\xic,\xiQ)= (e^{\ri\lambda}\xic h^{\dag},e^{\ri\theta}\xiQ h^{\dag})$, where $h\in (\mathbb{Z}_2)_{\mathrm{C}+\mathrm{Q}}$.
The $(\mathbb{Z}_2)_{\mathrm{C}+\mathrm{Q}}$ invariant combinations are $\xi_+=\xic\xiQ^\dag=\exp(\ri\varphi_+)$ and $\xi_-=(\xic\xiQ)^\dag=\exp(\ri\varphi_-)$,
i.e., $ \varphi_\pm = -\varphi_{\mathrm{Q}}\pm \varphi_\mathrm{C}$ correspond to the degrees of freedom above.
Noting $K=2$, and $a=\ri\xic\dd\xic=\dd{\varphi_\mathrm{C}}$, the symmetry generators are
\begin{align}
    U(e^{\ri\theta},C)&=\exp(\ri\frac{\theta}{2\pi}2\int_C\dd\varphi_{\mathrm{Q}}),\\
    W(n,C)&=\exp(\ri n\int_C\dd\varphi_\mathrm{C}),
\end{align}
where $n\in\mathbb{Z}$.

Let us now consider the second example, three-flavor QCD.
To simplify the situation, we assume that flavor masses are equal, and neglect the effect of chiral symmetry breaking, which does not affect our argument. The symmetry faithfully acting on quark fields is 
\begin{align}
    G=\frac{\mathrm{SU}(3)_\mathrm{C} \times \mathrm{SU}(3)_\mathrm{F} \times \mathrm{U}(1)_\mathrm{Q}}{(\mathbb{Z}_3)_{\mathrm{C}+\mathrm{Q}}\times (\mathbb{Z}_3)_{\mathrm{F}+\mathrm{Q}}}.
\end{align}
Here, the subscript $\mathrm{F}$ represents ``flavor''.
Since the center of color, flavor, and quark number share the same action on quarks, two of them are redundant.

In the CFL phase, a diquark field,
\begin{align}
    \Phi_{c_1 f_1}=\epsilon_{c_1c_2c_3}\epsilon_{f_1 f_2 f_3}(\psi^T_{c_2 f_2}\ri\gamma_0 \gamma_2 \gamma_5 \psi_{c_3 f_3})
\end{align}
is condensed. Here, $\epsilon_{abc}$ is the totally antisymmetric tensor,
$\gamma_i$ are the gamma matrices, and $\psi_{cf}$ is the quark field with flavor $f$ and color $c$ indices.
We consider the condensate $\expval{\Phi_{c_1f_1}}\propto \delta_{c_1f_1}$, which
 is invariant under simultaneous rotation of color and flavor; thus, it is called the color-flavor locking~\cite{Alford:1998mk}.
 The diquark has charge two under $\mathrm{U}(1)_\mathrm{Q}$, which is invariant under $e^{\ri\pi}\in(\mathrm{Z}_2)_\mathrm{Q}\subset \mathrm{U}(1)_\mathrm{Q}$.
In the end, the symmetry $G$ is broken into
\begin{align}
    H=
    \frac{\mathrm{SU}(3)_{\mathrm{C}+\mathrm{F}} \times  (\mathbb{Z}_{2})_\mathrm{Q}}{(\mathbb{Z}_3)_{\mathrm{C}+\mathrm{F}}}.
\end{align}
 The coset space is
\begin{align}
    G/H\simeq \frac{\mathrm{SU}(3)_{\mathrm{C}-\mathrm{F}}\times \mathrm{U}(1)_\mathrm{Q}}{(\mathbb{Z}_3)_{\mathrm{C}-\mathrm{F}}\times(\mathbb{Z}_2)_\mathrm{Q}}
    \simeq \frac{\mathrm{SU}(3)_{\mathrm{C}-\mathrm{F}}\times \mathrm{U}(1)_{q}}{(\mathbb{Z}_3)_{\mathrm{C}-\mathrm{F}}},
\end{align}
where $\mathrm{U}(1)_q=\mathrm{U}(1)/\mathbb{Z}_2$.
 Taking $(\xic,\xi_q)\in (\mathrm{SU}(3)_{\mathrm{C}-\mathrm{F}} \times \mathrm{U}(1)_{q})/(\mathbb{Z}_3)_{\mathrm{C}-\mathrm{F}}$, we can construct topological and gauge invariant objects,
\begin{align}
    U(e^{\ri\theta},C)&=\exp(\ri\frac{\theta}{2\pi}3\int_C \dd{\varphi_q}),\notag\\
    W(\rho,C)&=\frac{1}{d_\rho}\tr\rho\qty[\text{P}\exp(\ri\int_C a)],
\end{align}
which are generators of $\mathrm{U}(1)$ and $\mathbb{Z}_3$ two-form symmetries, respectively. Here, $a=\ri \xic\dd\xic^\dag$, and $\dd\varphi_q=\ri\xi_q \dd\xi_q^\dag$.
Their charged objects are worldsurfaces of abelian and nonabelian vortices.

\paragraph{Winding v.s. holonomy}
We have shown the emergent $\mathbb{Z}_K$ $(d-2)$-form symmetry in Higgs phases with superfluidity. In the previous works,
the holonomy of the gauge group, the Wilson loop, is employed to characterize the Higgs phases~\cite{Cherman:2018jir,Hirono:2018fjr,Hirono:2019oup,Cherman:2020hbe}. Let us discuss the relation between our symmetry generator and the Wilson loop.
The gauge invariance implies that the effective Lagrangian is 
a function of $(a_\mu-A_\mu)$ with $a_\mu =\ri \xic\partial_\mu\xic^\dag$. In the existence of a vortex defect,
if the minimum-energy solution is given by $a_\mu=A_\mu$ on a large closed line $C$,
\begin{equation}
        \expval{\tr\rho\qty[\mathrm{P}\exp(\ri \int_C A)]}=\expval{\tr\rho\qty[\mathrm{P}\exp(\ri \int_C a)]}
    \label{eq:winding_holonomy_relation}
\end{equation}
holds. That is, a winding phase and the Wilson loop give the same value. This is the case for the CFL phase with a common mass of quarks, where $\mathbb{Z}_3$ two-form symmetry emerges in which the symmetry generator is the Wilson loop~\cite{Hirono:2018fjr}. We note that, in general, if there is a quark mass difference, $a_\mu\neq A_\mu$, and Eq.~\eqref{eq:winding_holonomy_relation} does not hold. In this case, the Wilson loop is no longer $\mathbb{Z}_3$ valued. Nevertheless, the winding phase still is $\mathbb{Z}_3$ valued as long as the coset space does not change. This is the advantage of our formulation.

\paragraph{Phase transition v.s. continuity}
The possibility of the phase transition using holonomy was discussed in Ref.~\cite{Cherman:2018jir,Cherman:2020hbe}.
The existence and absence of nontrivial holonomy correspond to the existence and absence of emergent $\mathbb{Z}_K$ symmetry in our formalism.
Here, we would like to discuss the phase transition from a different point of view.
Roughly speaking, the emergent $\mathbb{Z}_K$ symmetry detects the different symmetry breaking patterns: $\mathrm{U}(1)_\mathrm{Q}\to \mathbb{Z}_q$ and $\mathrm{U}(1)_\mathrm{B}=\mathrm{U}(1)_\mathrm{Q}/\mathbb{Z}_K\to \mathbb{Z}_q$.
Although the breaking pattern $\mathrm{U}(1)_\mathrm{Q}\to \mathbb{Z}_q$ is not precise as the global symmetry breaking because of redundancy,
we propose the following gauge-invariant order parameter,
\begin{equation}
    \Delta\coloneqq\lim_{|x-y|\to\infty}\expval{\Phi^\dag(y)\rho_\Phi\qty[\mathrm{P}\exp(\ri \int_x^y A)] \Phi(x)},
    \label{eq:Higgs_oder_parameter}
\end{equation}
where $\Phi$ has the representation $(\rho_{\Phi},q)$~\footnote{A similar order parameter is proposed to distinguish the confinement and Higgs phases in Refs.~\cite{Greensite:2021fyi}.}.
In the Higgs phase where fluctuations of gauge fields are suppressed, $\Delta\sim \expval*{\Phi^\dag(y)}\expval*{\Phi(x)}$; thus, $\Delta$ can be thought of as the order parameter of $\mathrm{U}(1)_\mathrm{Q}$ symmetry breaking. On the other hand, $\Delta$ will vanish in the confinement phase where $\mathrm{U}(1)_\mathrm{B}$ is broken, but $\mathrm{U}(1)_\mathrm{Q}$ does not.
Therefore, we expect the phase transition between the confinement and Higgs phases characterized by $\Delta$.
To confirm the conjecture of the phase transition, it is necessary to show nonperturbatively that the $\Delta$ vanishes in the confinement phase; we leave this for future work.

\acknowledgments
Y.H. thanks Y. Tanizaki for the useful discussion.
Y.H. is supported in part by JSPS KAKENHI Grant Number~21H0108.

\bibliographystyle{utphys}
\bibliography{vortex}

\end{document}